\documentclass[aps,prl,reprint,superscriptaddress,floatfix]{revtex4-2}

\usepackage{amsmath,amssymb}
\usepackage{bm}
\usepackage{graphicx}
\usepackage[T1]{fontenc}
\usepackage{hyperref}
\hypersetup{colorlinks=true,breaklinks=true,linkcolor={[rgb]{0.10,0.25,0.55}},citecolor={[rgb]{0.10,0.25,0.55}},urlcolor={[rgb]{0.10,0.25,0.55}}}

\providecommand{\K}{}\renewcommand{\K}{\mathbf{K}}
\providecommand{\qv}{}\renewcommand{\qv}{\mathbf{q}}
\providecommand{\rvec}{}\renewcommand{\rvec}{\mathbf{r}}
\providecommand{\gv}{}\renewcommand{\gv}{\mathbf{g}}
\providecommand{\Mmat}{}\renewcommand{\Mmat}{\mathbf{M}}
\providecommand{\Pq}{}\renewcommand{\Pq}{\mathbf{P}_{q}}
\providecommand{\khat}{}\renewcommand{\khat}{\hat{\mathbf{k}}}
\providecommand{\abaralpha}{}\renewcommand{\abaralpha}{\bar{\alpha}}
\providecommand{\ULDM}{}\renewcommand{\ULDM}{\mathrm{ULDM}}
\providecommand{\STF}{}\renewcommand{\STF}{\mathrm{STF}}
\providecommand{\qabs}[1]{}\renewcommand{\qabs}[1]{\left|#1\right|}

\begin{document}

\title{Rank-One Ultralight Dark Matter Response in a Full-Tensor Gravity Gradiometer}

\author{John J. Oh}
\email{johnoh@nims.re.kr}
\affiliation{National Institute for Mathematical Sciences, Daejeon 34047, Republic of Korea}

\author{Ho Jung Paik}
\email{hpaik@umd.edu}
\affiliation{Department of Physics, University of Maryland, College Park, MD 20742, USA}

\begin{abstract}
A full-tensor superconducting gravity gradiometer whose four test masses carry two compositions in a $2+2$ pattern acquires a composition-dependent channel for dilaton-coupled ultralight dark matter. The reconstructed response is the rank-one tensor $\K=\gv_\phi(t)\otimes\qv$, with the axis of $\qv$ fixed in advance by the material assignment, and obeys the parameter-free null condition $\K(\mathbf{I}-\Pq)=0$, which no gravity gradient or platform rotation alone can satisfy; it vanishes for a single composition. For copper and lead, the sensitivity reaches $d_{\hat m}-d_g\simeq3.4\times10^{-8}$, and $4.9\times10^{-10}$ with upgraded accelerometers.
\end{abstract}

\maketitle

If dark matter is a scalar field of low mass, its local realization in the Galactic halo is a classical wave: the field oscillates coherently at the Compton frequency $f_\phi=m_\phi c^2/h$, with a coherence time set by the velocity dispersion of the halo~\cite{HuBarkana2000,HuiOstriker2017,FreeseLisantiSavage2013}. A field that couples to the Standard Model through a dilaton-like interaction varies the fundamental constants that set atomic and nuclear binding energies, and the variation is composition dependent. Two bodies of different composition acquire accelerations that differ in proportion to their dilaton charges~\cite{DamourDonoghue2010,DamourDonoghueCQG2010,Hees2018}. Static tests of the universality of free fall probe the same effect, and already constrain such couplings. Differential accelerometers, torsion balances and dual-species atom interferometers have been proposed as oscillating search platforms~\cite{Touboul2022,Berge2018,Graham2016}.

Superconducting gravity gradiometers (SGGs) with levitated or suspended test masses~(TMs) were developed for geodesy and, more recently, as full-tensor gravitational-wave detectors~\cite{MoodyChanPaik1986,ChanMoodyPaik1987,Paik2016,Paik2020,Griggs2017}. The tetrahedral form is used in the ground-based mSOGRO design and in the proposed lunar LANGO observatory~\cite{LANGO2026,Metzler2022}. Four TMs at the vertices reconstruct the full acceleration-gradient tensor, from which a gravitational wave is read out. Such an instrument is, in its usual operation, blind to dark matter. With a single composition, the equivalence-principle-violating acceleration is common to all four masses and is rejected with every other common-mode disturbance.

\begin{figure}[t]
\centering
\includegraphics[width=0.8\columnwidth]{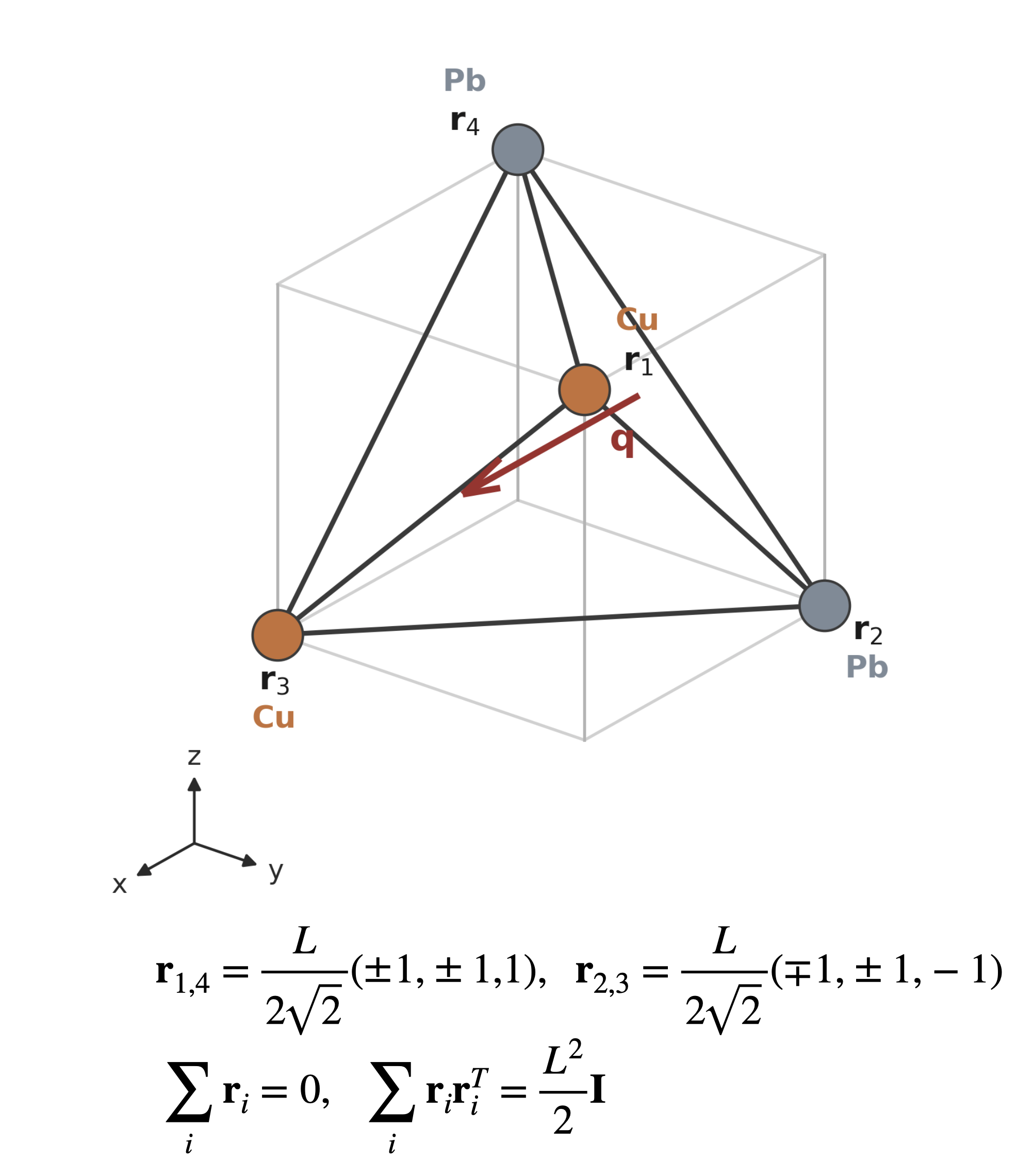}
\caption{\label{fig:geom}A $2+2$ composition assignment on the regular tetrahedron of Eq.~(\ref{eq:vertices}), copper on $\{\rvec_1,\rvec_3\}$ and lead on $\{\rvec_2,\rvec_4\}$, which puts $\qv$ along $\hat{\bm x}$; the other two splits select $\hat{\bm y}$ and $\hat{\bm z}$ and are equivalent to it.}
\end{figure}

\begin{figure}[t]
\centering
\includegraphics[width=\columnwidth]{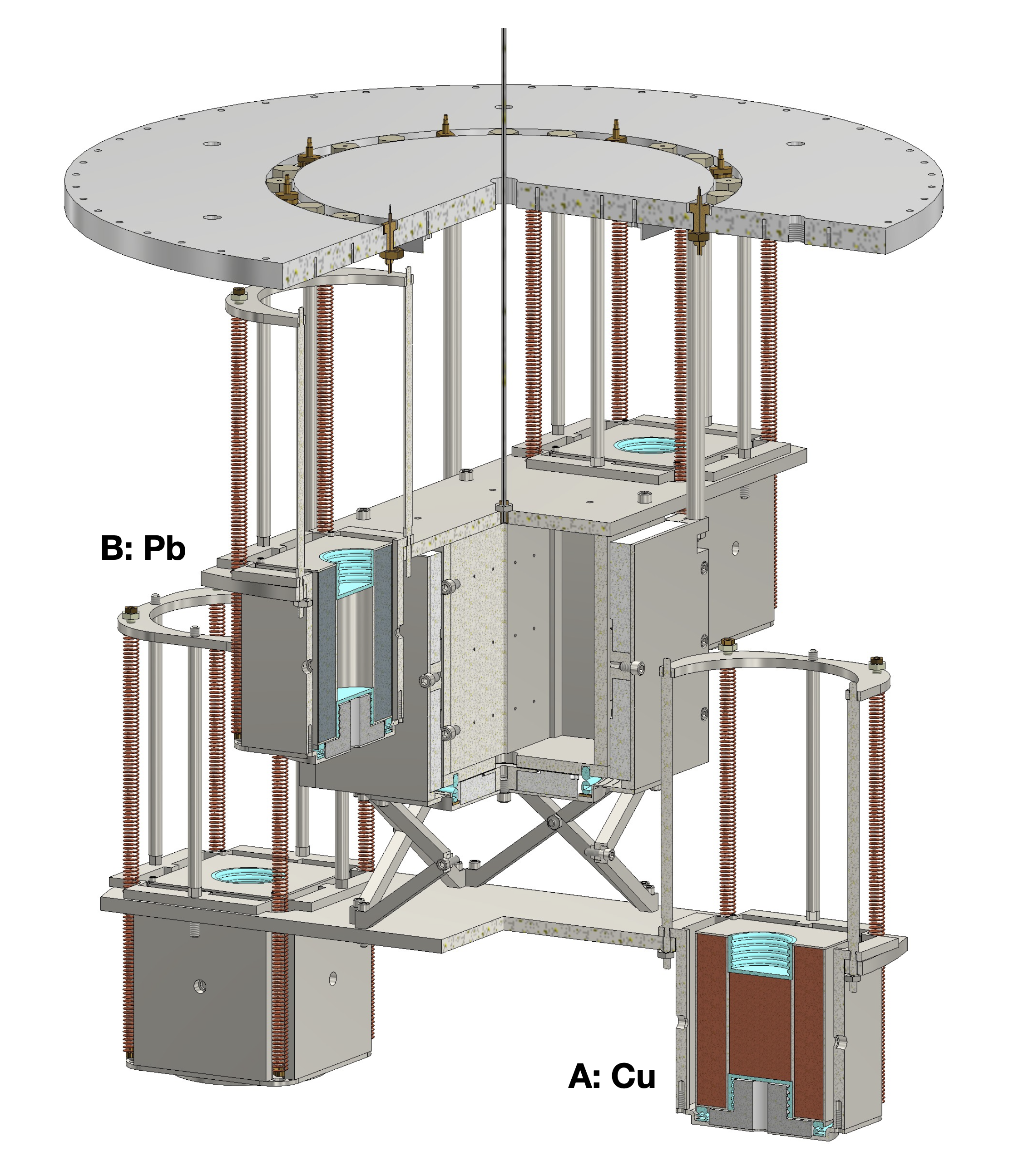}
\caption{\label{fig:mech}The assignment of Fig.~\ref{fig:geom} on the mSOGRO mechanical design; only the choice of TM material changes. The whole platform hangs as a pendulum, which isolates it from the angular motion of the ground.}
\end{figure}

We show that the same hardware becomes sensitive to an oscillating scalar field once TMs of more than one composition are introduced, and that the spatial arrangement of these compositions determines the detector response. Assigning two materials to the four vertices in a $2+2$ pattern (Figs.~\ref{fig:geom} and~\ref{fig:mech}) turns $\Delta\abaralpha_{AB}$ into a time-independent vector $\qv$ in the reconstructed tensor: the dark-matter response becomes the rank-one tensor $\K_{\ULDM}(t)=\gv_\phi(t)\otimes\qv$, whose field gradient vector is stochastic and whose composition vector has an axis fixed by the material assignment and the geometry. No dark-matter sensor is added and no dark-matter-specific readout is built; the signal appears in the channels that already carry gravitational waves, distinguished from them by a null condition that follows from the factorization. The vector $\qv$ is real, and the axis it defines is fixed once the materials are assigned; only its sense depends on the sign of $\Delta\abaralpha_{AB}$, which carries the unknown couplings. The reconstructed tensor is therefore confined at every instant to the matrices whose rows are multiples of $\qv^{\rm T}$,
\begin{equation}
\K_{\ULDM}(\mathbf{I}-\Pq)=0,
\qquad
\Pq=\frac{\qv\,\qv^{\rm T}}{\qabs{\qv}^{2}} .
\label{eq:null}
\end{equation}
$\mathbf{I}-\Pq$ annihilates $\qv$, so Eq.~(\ref{eq:null}) states that the response has no component along the two directions orthogonal to $\qv$. A gravity gradient in vacuum cannot satisfy it. Symmetry forces a rank-one tensor whose rows are multiples of $\qv^{\rm T}$ to be proportional to $\qv\otimes\qv$, and tracelessness then forces it to vanish. A nonzero Euler term cannot satisfy it either, because a real antisymmetric $3\times3$ matrix has rank two. The Coriolis term leaves a tensor that need not be antisymmetric; for a rigid platform its rank-one part has right factor along the Earth's axis, not along $\qv$, and the companion paper treats it~\cite{companion}. Each is excluded on its own, though a correlated sum of the two is not. What can satisfy it in general is a disturbance whose differential force follows the same material assignment. A companion paper~\cite{companion} gives the full derivation, the corrections from working in a rotating frame, and the implementation of the vertical readout.

\textit{The signal.}---A real scalar $\phi$ of mass $m_\phi$ at the solar position is strongly non-relativistic for a virial velocity $v_{\rm DM}\simeq10^{-3}c$, and oscillates at
\begin{equation}
f_\phi=\frac{m_\phi c^2}{h}
\simeq 2.4\times10^{2}\,\mathrm{Hz}\left(\frac{m_\phi c^2}{10^{-12}\,\mathrm{eV}}\right),
\label{eq:fphi}
\end{equation}
placing the mass window $m_\phi c^2=10^{-18}$--$10^{-15}\,\mathrm{eV}$ at $f_\phi\simeq0.24\,\mathrm{mHz}$--$0.24\,\mathrm{Hz}$, within the band of a terrestrial gradiometer. The de Broglie wavelength exceeds any laboratory baseline by more than eleven orders of magnitude, $\lvert\bm k_\phi\rvert L\lesssim2\times10^{-12}$ across the mass window, so the four TMs see one and the same field gradient. A differential signal cannot then arise from a phase difference across the array; it can arise only from the masses made of different materials. The occupation of a de Broglie volume is $10^{66}$ or larger, which allows the field to be treated as a classical wave rather than as a quantum operator~\cite{HuBarkana2000,HuiOstriker2017}. The halo has a velocity spread, so modes moving at slightly different speeds oscillate at slightly different frequencies and dephase. The fractional spread is $\Delta\omega_\phi/\omega_\phi\sim v_{\rm DM}^{2}\sim10^{-6}$, which means the field holds a definite amplitude and phase for $Q_{\rm coh}=\omega_\phi/\Delta\omega_\phi\sim10^{6}$ cycles, a coherence time $\tau_{\rm coh}\sim Q_{\rm coh}/f_\phi$. The signal is therefore a line of fractional width $10^{-6}$ everywhere in the band, far narrower than any broadband background.

A dilaton-like scalar couples to the electromagnetic, gluonic and fermion-mass sectors of the Standard Model, and the force on a macroscopic TM enters not through that coupling directly but through the field dependence of the rest mass. Let $\alpha_A=\partial\ln m_A(\varphi)/\partial\varphi$ be the scalar charge of body $A$, with the field normalized as $\varphi=\kappa\phi$, $\kappa=\sqrt{4\pi G}$. The composition enters only through $\alpha_A$, built from the dilaton charges of Damour and Donoghue~\cite{DamourDonoghue2010,Hees2018} tabulated in the End Matter (Table~\ref{tab:charges}). The acceleration of $A$ is then
\begin{equation}
\bm a_A^\varphi(t)=-\kappa\,\alpha_A\,\nabla\phi(t)\equiv\alpha_A\,\gv_\phi(t),
\label{eq:aA}
\end{equation}
so that every TM is pushed along one common, acceleration-valued field $\gv_\phi$ with a strength set by its own charge. Its amplitude follows from the local density and the halo speed (End Matter),
\begin{equation}
A_0\equiv\kappa\lvert\nabla\phi\rvert_{\rm amp}=\sqrt{8\pi G\rho_{\rm DM}}\;v_{\rm DM}=2.95\times10^{-10}\,\mathrm{m\,s^{-2}} ,
\label{eq:A0}
\end{equation}
with $\hbar=c=1$, for $\rho_{\rm DM}=0.4\,\mathrm{GeV}/c^2\,\mathrm{cm^{-3}}$~\cite{Read2014} and rms speed $v_{\rm DM}=\sqrt{3/2}\,v_0=269\,\mathrm{km\,s^{-1}}$ at $v_0=220\,\mathrm{km\,s^{-1}}$~\cite{FreeseLisantiSavage2013}. The scalar mass cancels between the amplitude and the momentum, so $A_0$ depends only on the halo, not on $m_\phi$, the TM or the baseline. The composition enters through $\alpha_A$ instead. A single plane wave is an idealization: within one coherence interval a virialized halo gives $\gv_\phi(t)=\bm u\cos\omega_\phi t+\bm v\sin\omega_\phi t$ with $\bm u$ and $\bm v$ in general not parallel, so the gradient traces an ellipse~\cite{LisantiMoschellaTerrano2021,Centers2021}. The charge $\alpha_A$ is defined for a single nuclide, whereas a TM could be an alloy, so a rule is needed for combining constituents. The acceleration of the center of mass weights each by its own mass, so the relevant quantity is the mass-weighted mean charge $\abaralpha_A$, and the differential acceleration between two materials is
\begin{equation}
\Delta\bm a_{AB}^\varphi(t)=\Delta\abaralpha_{AB}\,\gv_\phi(t),
\qquad
\Delta\abaralpha_{AB}\equiv\abaralpha_A-\abaralpha_B .
\label{eq:diffacc}
\end{equation}

\textit{Tetrahedral reconstruction.}---The four TMs are placed at the vertices of a regular tetrahedron,
\begin{equation}
\begin{aligned}
\rvec_1&=\tfrac{L}{2\sqrt2}(1,1,1)^{\rm T}, & \rvec_2&=\tfrac{L}{2\sqrt2}(-1,1,-1)^{\rm T},\\ \rvec_3&=\tfrac{L}{2\sqrt2}(1,-1,-1)^{\rm T}, & \rvec_4&=\tfrac{L}{2\sqrt2}(-1,-1,1)^{\rm T},
\end{aligned}
\label{eq:vertices}
\end{equation}
which are centered, $\sum_i\rvec_i=0$, with isotropic second-moment matrix $\Mmat=\sum_i\rvec_i\rvec_i^{\rm T}=\tfrac{L^2}{2}\,\mathbf I$; the vertices occupy alternate corners of a cube of side $L/\sqrt2$ and all six pairs are separated by $L$. The array is small compared with the scale on which the acceleration varies, so at each mass the acceleration is a constant plus a term linear in position, $\bm a_i(t)=\bm A(t)+\K(t)\rvec_i$, with $\bm A=\tfrac14\sum_i\bm a_i$ the common translational mode. Multiplying on the right by $\rvec_i^{\rm T}$, summing and multiplying by $\Mmat^{-1}$ gives
\begin{equation}
\K=\left[\sum_{i=1}^4\bm a_i\rvec_i^{\rm T}\right]\Mmat^{-1} ,
\label{eq:Kreco}
\end{equation}
the common mode dropping out of the sum because $\sum_i\rvec_i^{\rm T}=0$.
A Newtonian gradient $\Gamma_{ij}=-\partial_i\partial_j\Phi$ and the geodesic-deviation term of a gravitational wave are symmetric and traceless in vacuum, so both reside in the symmetric trace-free (STF) sector. The antisymmetric sector comes from rotation of the platform, whose Euler term measures the angular acceleration. A mass hung on a wire is stiff in the vertical, and that mode can be softened by a mechanical spring combined with a superconducting negative spring, a technique already demonstrated in this instrument family~\cite{Parke1984,Paik1996}. Each mass is then read in all three directions, so all nine components of $\K$ are formed, the tilts are measured rather than assumed small, and the trace is available~\cite{companion}.

\textit{The rank-one response.}---Each TM is pushed along the common field gradient with a strength set by its own charge, so by Eq.~(\ref{eq:aA})
$\bm a_i^{\ULDM}(t)=\abaralpha_i\,\gv_\phi(t)$.
Inserting this into Eq.~(\ref{eq:Kreco}) and using that $\gv_\phi$ does not depend on $i$ while $\Mmat^{-1}$ is symmetric, the sum factorizes. Defining the \emph{composition vector}
\begin{equation}
\qv\equiv\Mmat^{-1}\sum_{i=1}^{4}\abaralpha_i\rvec_i ,
\label{eq:qdef}
\end{equation}
the response is the outer product
\begin{equation}
\K_{\ULDM}(t)=\gv_\phi(t)\otimes\qv ,
\label{eq:Kuldm}
\end{equation}
a tensor of rank one at every instant. The factorization is exact in this
common-gradient limit, finite-wavelength corrections entering at
$O(\lvert\bm k_\phi\rvert L)$. The field gradient vector carries the entire time dependence, and the composition vector the entire composition dependence. For a single plane-wave mode, this reduces to
\begin{equation}
\K_{\ULDM}(t)=-A_0\cos(\omega_\phi t+\theta)\,\khat\otimes\qv .
\label{eq:Kuldm-single}
\end{equation}

Writing $\abaralpha_i=\abaralpha_0+\delta\abaralpha_i$ with $\abaralpha_0$ the mean charge, the mean part drops out of Eq.~(\ref{eq:qdef}) because the vertices are centered. The mean charge appears instead in the translational common mode $\bm A_{\ULDM}=\abaralpha_0\gv_\phi$, which any gradiometer rejects along with platform acceleration. Only the deviations survive, and in particular for TMs of a single composition every $\delta\abaralpha_i$ vanishes and
$\qv=0$
identically. This is why an existing single-composition array carries no dark-matter channel, and why supplying one requires no dark-matter-specific sensor or readout.

The outer product $\gv_\phi\otimes\qv$ is neither symmetric nor antisymmetric, nor is it traceless, so it fills all three sectors of the tensor. Their squared magnitudes are
\begin{equation}
\begin{aligned}
\lVert\K_{\rm tr}\rVert^2&=\tfrac13(\gv_\phi\!\cdot\!\qv)^2,\\
\lVert\K_{\rm A}\rVert^2&=\tfrac12\lvert\gv_\phi\!\times\!\qv\rvert^2,\\
\lVert\K_{\STF}\rVert^2&=\tfrac12\lvert\gv_\phi\rvert^2\lvert\qv\rvert^2+\tfrac16(\gv_\phi\!\cdot\!\qv)^2,
\end{aligned}
\label{eq:norms}
\end{equation}
and they sum to $\lvert\gv_\phi\rvert^2\lvert\qv\rvert^2$. These are algebraic identities, true at every instant for any $\gv_\phi$, so the three are locked to one another and none is a free parameter.

Gravitational tidal fields in vacuum leave the trace empty. A gravity gradient in vacuum is traceless because the potential obeys Laplace's equation, and a gravitational wave is traceless in general relativity, whereas
\begin{equation}
\operatorname{Tr}\K_{\ULDM}=\gv_\phi\!\cdot\!\qv
\label{eq:trace}
\end{equation}
does not vanish. The dark matter therefore enters a channel that neither background occupies. A nonzero trace does not violate Laplace's equation. $\K_{\ULDM}$ is not the gradient of an acceleration field but an effective tensor produced by four masses responding with different strengths to one common field.

\textit{The null condition.}---Every row of $\gv_\phi\otimes\qv$ is a multiple of $\qv^{\rm T}$, so the tensor annihilates every direction orthogonal to $\qv$, as shown by Eq.~(\ref{eq:null}). Since $\mathbf{I}-\Pq$ has rank two, the equation imposes two conditions per row, six in all. The axis of $\qv$ is fixed before any data are taken, and $\Pq$ depends only on that axis; the sense is a labeling convention, taken here as $\Delta\abaralpha_{AB}>0$, so the unknown couplings in $\qabs{\qv}\propto\Delta\abaralpha_{AB}$ never enter. What it tests is that every row is a multiple of $\qv^{\rm T}$. A rank-one tensor pointing any other way fails. Because the condition holds at each instant for any $\gv_\phi$, it also holds in a virialized halo, where the rows no longer share a phase but each remains a multiple of $\qv^{\rm T}$~\cite{companion}.

A disturbance passes this test whenever it has the same structure as the signal. Magnetic differential acceleration is a case that matters. The two materials differ in susceptibility, so a magnetic force acts on them differently and produces a rank-one tensor with the same $\qv$. The magnetic response even reverses when the materials are exchanged, just as the signal does, so a geometric check cannot separate the two.

The test applies to the candidate component estimated with the gradient and rotation as nuisance processes~\cite{companion}. A candidate that fails under every admissible background is excluded as dark matter. Passing does not establish dark matter. Confirmation requires two further steps. The first is a set of physical controls: magnetic shielding, continuous magnetometer monitoring of the residual field during the run, and deliberate injection of a known field gradient~\cite{companion}. The second is a way to change the composition while leaving the hardware alone.

Two tetrahedra interleaved on one frame provide the second. The cryostat can also be mounted on a turntable and rotated by $90$ or $180$ degrees, which moves the composition axis in the laboratory while leaving the gravity gradient and the magnetic environment as they are. A second array can be built, in alternative realizations, with $-\qv$, with $\qv$ along the orthogonal horizontal axis, or with a single composition, giving respectively the material exchange, an independent axis, and the $\qv=0$ control.

\textit{Two-material assignment.}---There are six $2+2$ assignments of the two materials among the four TMs, falling into three pairs that select the three cube axes. Since $\Mmat$ is isotropic, the three are equivalent, so the composition axis is free. Exchanging the two materials within a pair reverses the sign of $\qv$ and changes nothing else. Taking material $A$ on one pair of opposite vertices and $B$ on the other two, each group sum lies along one cube axis and Eq.~(\ref{eq:qdef}) gives
\begin{equation}
\qv=\frac{\sqrt2\,\Delta\abaralpha_{AB}}{L}\,\hat{\bm x},
\qquad
\qabs{\qv}=\frac{\sqrt2\,\lvert\Delta\abaralpha_{AB}\rvert}{L} .
\label{eq:qresult}
\end{equation} 

\begin{figure}[t]
\centering
\includegraphics[width=\columnwidth]{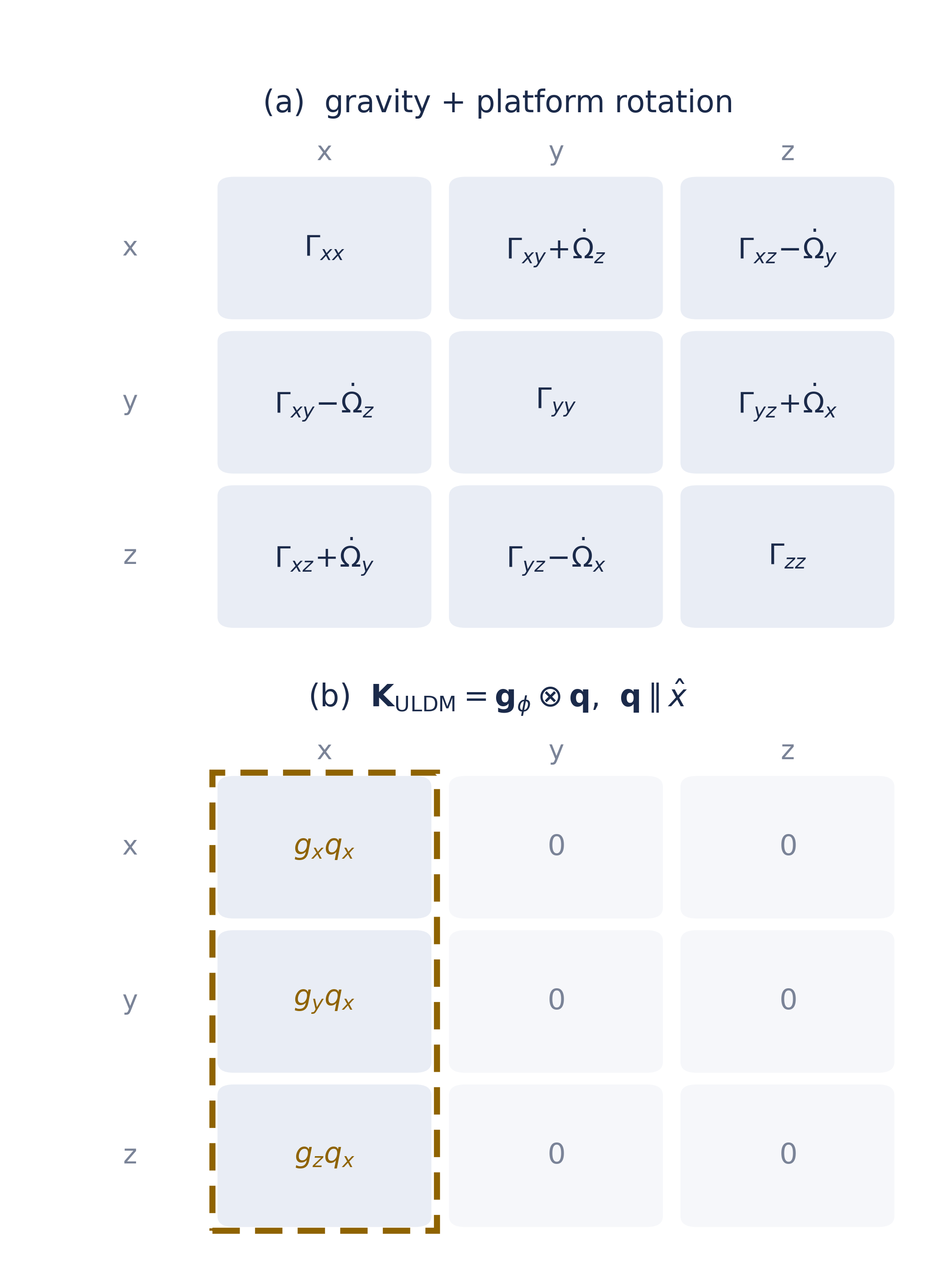}
\caption{\label{fig:channels}Occupation of the nine tensor components. (a) Gravity fills the symmetric part and platform rotation the antisymmetric part; with all three rows read, the symmetric and antisymmetric combinations of the $xz$ and $yz$ pairs separate the gradient from the tilt, as the $xy$ pair already does for the yaw. (b) The dark-matter response is $\gv_\phi\otimes\qv$. For the assignment of Fig.~\ref{fig:geom}, which puts $\qv$ along $\hat{\bm x}$, only the first column survives; the tensor annihilates every direction orthogonal to $\qv$, whatever the direction of the gradient. The other two splits put $\qv$ along $\hat{\bm y}$ and $\hat{\bm z}$ and fill the second and third columns instead.}
\end{figure}

With $\qv$ along $\hat{\bm x}$, the response occupies the first column alone:
\begin{equation}
\K_{\ULDM}=q_x\begin{pmatrix}g_x&0&0\\ g_y&0&0\\ g_z&0&0\end{pmatrix},
\label{eq:signalpattern}
\end{equation}
so nine channels are fixed by three numbers (Fig.~\ref{fig:channels}) and Eq.~(\ref{eq:null}) imposes six
conditions, all of them now testable. The trace is the one channel that gravity
leaves empty,
$\operatorname{Tr}\K_{\rm meas}=g_xq_x+\operatorname{Tr}\mathbf n$,
because a gravity gradient in vacuum obeys $\operatorname{Tr}\bm\Gamma=-4\pi G\rho_{\rm
local}=0$, a gravitational wave is traceless, and the Euler term has no trace.
Outside the trace, the signal-bearing components stay mixed with a gravitational
or a rotational background, so a candidate must reproduce the whole three-row
pattern rather than part of it. The narrow linewidth, the fixed
axis of the rows, and the sidereal modulation provide complementary
evidence, and the companion paper develops the analysis~\cite{companion}.

\textit{Sensitivity.}---The signal is an acceleration, $a_{\rm sig}=A_0\Delta\abaralpha$. Its dependence on $\Delta\abaralpha$ is the violation of the equivalence principle: two materials fall differently.

For an integration time $\tau_{\rm int}$, what enters the signal-to-noise ratio is not $\tau_{\rm int}$ itself. While $\tau_{\rm int}\le\tau_{\rm coh}$ the field keeps a definite phase and $\tau_{\rm eff}=\tau_{\rm int}$. For longer observations the phase resets every $\tau_{\rm coh}$, the record breaks into segments, and averaging them leaves $\tau_{\rm eff}=\sqrt{\tau_{\rm coh}\tau_{\rm int}}$~\cite{Manley2021}. The coherence boundary lies at $f_\phi=Q_{\rm coh}/\tau_{\rm int}\simeq30\,$mHz for a year of data. The noise is the thermal acceleration noise of the differential mode of a mass pair, $S_a=8k_BT\omega_0/mQ$, twice the single-mass value~\cite{Paik1996}. The signal does not depend on $m$ while the noise amplitude falls as $1/\sqrt m$, so a heavier TM gains $\sqrt m$. The shell and the niobium hardware are common to the two TMs, so they cancel in the charge difference and enter only through the total mass; the core fraction $0.90$ of the Maryland-designed TMs multiplies $\Delta\abaralpha$, so the closed form below carries $\Delta Q'_{\hat m,\rm eff}=0.90\,\Delta Q'_{\hat m}$~\cite{companion}. Setting $\mathrm{SNR}=1$, as in comparable proposals~\cite{Manley2021}, and writing $\Delta\abaralpha=\sum_c d_c\Delta Q'_c$, the sensitivity in the couplings follows~\cite{companion},
\begin{equation}
d_c^{\rm max}\propto\frac{1}{\Delta Q'_c}
\sqrt{\frac{8T\omega_0}{3mQ}}\;\frac{1}{\sqrt{\tau_{\rm eff}}} .
\label{eq:sens}
\end{equation}
Above the coherence boundary $d_c^{\rm max}$ falls only as $\tau_{\rm int}^{-1/4}$. The array size cancels. Since $\qabs{\qv}\propto1/L$, the tensor scales as $\K_{\ULDM}\propto1/L$, while the differential acceleration it produces, of order $\K_{\ULDM}L$, does not. A gravitational wave behaves in the opposite way: its tidal tensor is set by the wave, so its differential acceleration grows with the array. A larger array lowers the equivalent gradient noise but not the coupling sensitivity, because the ULDM response falls with the same $1/L$.

\begin{figure}[t]
\centering
\includegraphics[width=\columnwidth]{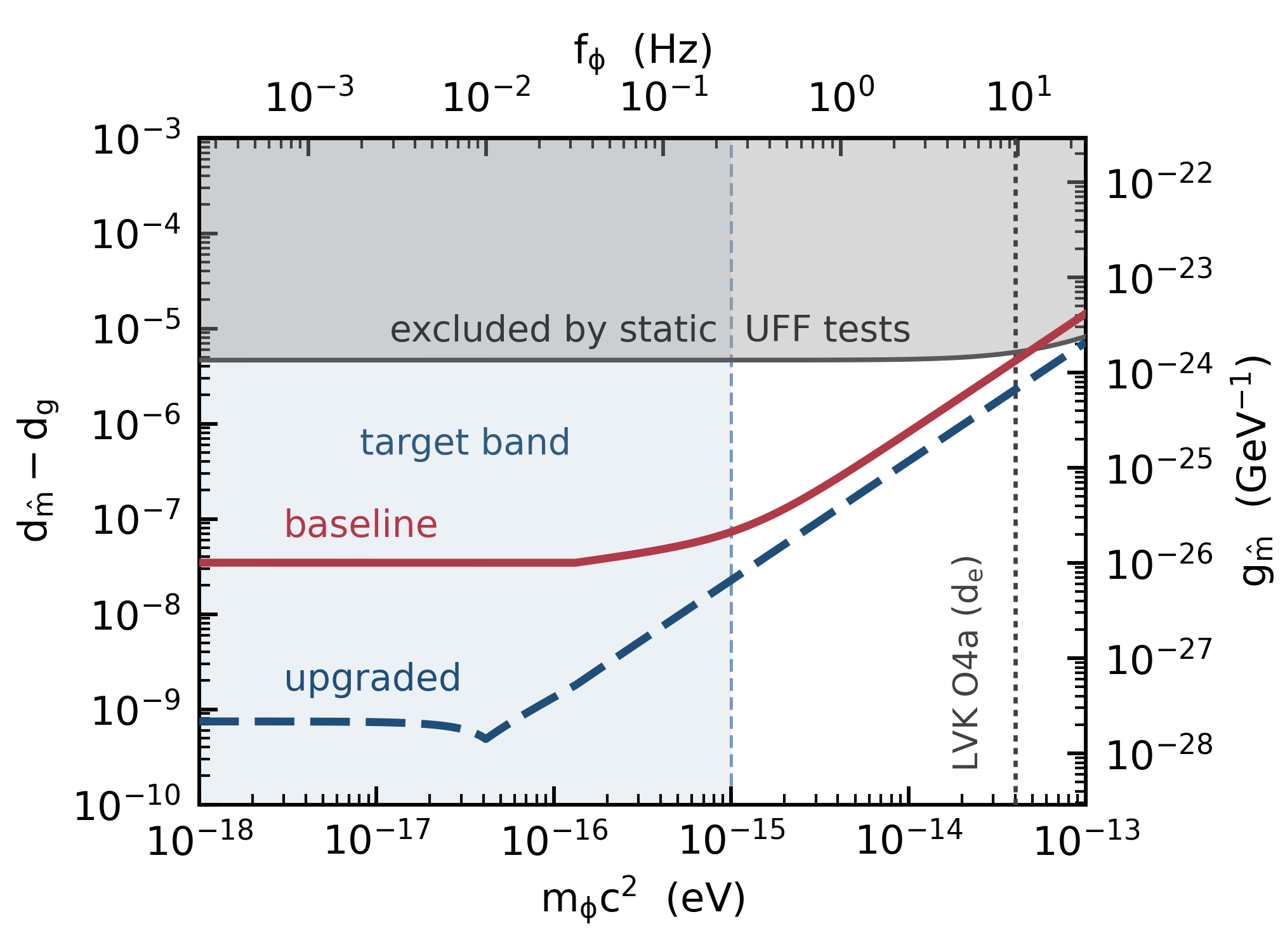}
\caption{\label{fig:sens}Instrument-noise-limited characteristic sensitivity in $d_{\hat m}-d_g$ versus ULDM mass, for the copper--lead pair in the $2+2$ assignment. The shaded region is excluded by static tests of the universality of free fall~\cite{Touboul2022,Schlamminger2008,Wagner2012}. The tinted band is the mass window of Eq.~(\ref{eq:fphi}), beyond which the curves are continued only to show where they meet the static bound; the dotted line gives the onset of the interferometer dilaton searches~\cite{LVKO4a2025}, which constrain photon and electron couplings, not the combination plotted here. Construction, parameters and caveats are given in the End Matter.}
\end{figure}

Figure~\ref{fig:sens} combines Eq.~(\ref{eq:sens}) with the SOGRO instrument noise model~\cite{Paik2016,Paik2020}, whose closed form is Eq.~(\ref{eq:noisemodel}) of the End Matter. In a year the baseline curve is lowest near $30\,$mHz, at $d_{\hat m}-d_g\simeq3.4\times10^{-8}$; upgrading the accelerometers to $m=40\,$kg, $T=0.1\,$K, $Q=10^{9}$ and $f_0=10\,$mHz lowers the curve and moves its minimum to $10\,$mHz, where it reaches $4.9\times10^{-10}$. The two minima differ by a factor of $70$, which comes entirely from parameters shared with the gravitational-wave channel.

\textit{Outlook.}---The vertical stiffness that has to be cancelled is $5.0\times10^{2}\,$N\,m$^{-1}$, about one third of the
stiffness already canceled in the Model-3 SGG~\cite{Paik1996}.  We propose to fully restore the vertical sensitivity in our ULDM apparatus. With the third row measured the null condition supplies six constraints, and a candidate signal must match a pattern in six channels rather than appear as an excess in one. This is not the only way to discriminate against the gravity gradient. Differencing two arrays cancels the common gradient~\cite{companion}.

Composition-dependent violation of the equivalence principle already underlies a mature class of dark-matter searches~\cite{Touboul2022,Berge2018,Graham2016}. Those instruments compare two materials along essentially a single axis and give one number. This array reconstructs the response as a tensor. A candidate is then characterized not only by its composition dependence but by which sector it occupies and in what proportion. The three norms obey the instantaneous algebraic relations of Eq.~(\ref{eq:norms}), and the amplitude in each turns with the Earth over a sidereal day. The measurement is of the same physical quantity; the discrimination available from it is not.

\textit{Data availability.}---The script that generates the sensitivity figure is archived at Zenodo~\cite{SensitivityCode2026}.

\begin{acknowledgments}
We thank J.~W. Lee for discussions on ultralight dark matter. JJO was supported by the Basic Science Research Program
through the National Research Foundation of Korea (NRF) funded by the Korea Ministry of Science and ICT (MSIT) (RS-2025-00555178) and the National Institute for Mathematical Sciences under the Primary Research Program (Project No. B26720000) supported by MSIT. HJP was supported by the National Science Foundation of the USA under grant PHY2207757.

The ideas, derivations, and proofs presented here are the authors' own. Claude Opus~5 and Fable~5 (Anthropic) were used to check selected calculations, to write the code that generates Fig.~\ref{fig:sens} from Eqs.~(\ref{eq:noisemodel}), (\ref{eq:sensclosed}) and (\ref{eq:microline}), and to assist with editing the manuscript, and GPT-5.5 and GPT-6.0 (OpenAI) were used as additional cross-checks. The authors directed all such use and independently verified all substantive outputs against their derivations and the cited literature. The authors take full responsibility for the content.
\end{acknowledgments}

\bibliographystyle{apsrev4-2}
\bibliography{sgg_uldm_theory}

@article{DamourDonoghue2010,
  author = {Damour, Thibault and Donoghue, John F.},
  title = {Equivalence Principle Violations and Couplings of a Light Dilaton},
  journal = {Physical Review D},
  volume = {82},
  pages = {084033},
  year = {2010},
  doi = {10.1103/PhysRevD.82.084033}
}

@article{Hees2018,
  author = {Hees, Aurélien and Minazzoli, Olivier and Savalle, Etienne and Stadnik, Yevgeny V. and Wolf, Peter},
  title = {Violation of the equivalence principle from light scalar dark matter},
  journal = {Physical Review D},
  volume = {98},
  pages = {064051},
  year = {2018},
  doi = {10.1103/PhysRevD.98.064051}
}

@article{Graham2016,
  author = {Graham, Peter W. and Kaplan, David E. and Mardon, Jeremy and Rajendran, Surjeet and Terrano, William A.},
  title = {Dark Matter Direct Detection with Accelerometers},
  journal = {Physical Review D},
  volume = {93},
  pages = {075029},
  year = {2016},
  doi = {10.1103/PhysRevD.93.075029}
}

@article{Berge2018,
  author = {Bergé, Joël and others},
  title = {MICROSCOPE Mission: First Constraints on the Violation of the Weak Equivalence Principle by a Light Scalar Dilaton},
  journal = {Physical Review Letters},
  volume = {120},
  pages = {141101},
  year = {2018},
  doi = {10.1103/PhysRevLett.120.141101}
}

@article{Touboul2022,
  author = {Touboul, Pierre and others},
  title = {MICROSCOPE Mission: Final Results of the Test of the Equivalence Principle},
  journal = {Physical Review Letters},
  volume = {129},
  pages = {121102},
  year = {2022},
  doi = {10.1103/PhysRevLett.129.121102}
}

@article{Metzler2022,
  author = {Metzler, Zachary and Collins, Christopher J. and Paik, Ho Jung and Shawhan, Peter S.},
  title = {Tetrahedral omnidirectional full-tensor gravitational wave detector},
  journal = {Classical and Quantum Gravity},
  volume = {39},
  pages = {225012},
  year = {2022},
  doi = {10.1088/1361-6382/ac902d}
}

@article{Paik2016,
  author = {Paik, Ho Jung and Griggs, Carl E. and Moody, Michael V. and Venkateswara, Krishna and Lee, Hyung Mok and Nielsen, Alex B. and Majorana, Ettore and Harms, Jan},
  title = {Low-frequency terrestrial tensor gravitational-wave detector},
  journal = {Classical and Quantum Gravity},
  volume = {33},
  pages = {075003},
  year = {2016},
  doi = {10.1088/0264-9381/33/7/075003}
}

@article{Paik2020,
  author = {Paik, Ho Jung and Moody, Michael V. and Norton, Ronald S.},
  title = {SOGRO---terrestrial full-tensor detector for mid-frequency gravitational waves},
  journal = {International Journal of Modern Physics D},
  volume = {29},
  pages = {1940001},
  year = {2020},
  doi = {10.1142/S0218271819400017}
}

@article{LANGO2026,
  author = {Paik, Ho Jung and Metzler, Zachary and Miller, M. Coleman and Collins, Christopher J. and Neal, Clive R. and Turyshev, Slava G.},
  title = {Lunar accelerometer network gravitational observatory (LANGO)},
  journal = {Classical and Quantum Gravity},
  volume = {43},
  pages = {095023},
  year = {2026},
  doi = {10.1088/1361-6382/ae676d}
}

@article{Read2014,
  author = {Read, Justin I.},
  title = {The local dark matter density},
  journal = {Journal of Physics G: Nuclear and Particle Physics},
  volume = {41},
  pages = {063101},
  year = {2014},
  doi = {10.1088/0954-3899/41/6/063101}
}

@article{FreeseLisantiSavage2013,
  author = {Freese, Katherine and Lisanti, Mariangela and Savage, Christopher},
  title = {Colloquium: Annual modulation of dark matter},
  journal = {Reviews of Modern Physics},
  volume = {85},
  pages = {1561},
  year = {2013},
  doi = {10.1103/RevModPhys.85.1561}
}

@article{HuBarkana2000,
  author = {Hu, Wayne and Barkana, Rennan and Gruzinov, Andrei},
  title = {Fuzzy Cold Dark Matter: The Wave Properties of Ultralight Particles},
  journal = {Physical Review Letters},
  volume = {85},
  pages = {1158},
  year = {2000},
  doi = {10.1103/PhysRevLett.85.1158}
}

@article{HuiOstriker2017,
  author = {Hui, Lam and Ostriker, Jeremiah P. and Tremaine, Scott and Witten, Edward},
  title = {Ultralight scalars as cosmological dark matter},
  journal = {Physical Review D},
  volume = {95},
  pages = {043541},
  year = {2017},
  doi = {10.1103/PhysRevD.95.043541}
}

@article{LisantiMoschellaTerrano2021,
  author = {Lisanti, Mariangela and Moschella, Matthew and Terrano, William},
  title = {Stochastic properties of ultralight scalar field gradients},
  journal = {Physical Review D},
  volume = {104},
  pages = {055037},
  year = {2021},
  doi = {10.1103/PhysRevD.104.055037}
}

@article{Centers2021,
  author = {Centers, Gary P. and others},
  title = {Stochastic fluctuations of bosonic dark matter},
  journal = {Nature Communications},
  volume = {12},
  pages = {7321},
  year = {2021},
  doi = {10.1038/s41467-021-27632-7}
}

@article{DamourDonoghueCQG2010,
  author = {Damour, Thibault and Donoghue, John F.},
  title = {Phenomenology of the equivalence principle with light scalars},
  journal = {Classical and Quantum Gravity},
  volume = {27},
  pages = {202001},
  year = {2010},
  doi = {10.1088/0264-9381/27/20/202001}
}

@inproceedings{Parke1984,
  author = {Parke, J. W. and others},
  title = {A superconducting negative spring},
  booktitle = {Proceedings of the 10th International Cryogenic Engineering Conference},
  editor = {Collan, D. H. and others},
  publisher = {Butterworth},
  address = {Surrey},
  pages = {361--364},
  year = {1984}
}

@article{MoodyChanPaik1986,
  author = {Moody, M. V. and Chan, H. A. and Paik, H. J.},
  title = {Superconducting gravity gradiometer for space and terrestrial applications},
  journal = {Journal of Applied Physics},
  volume = {60},
  pages = {4308--4315},
  year = {1986},
  doi = {10.1063/1.337474}
}

@article{ChanMoodyPaik1987,
  author = {Chan, H. A. and Moody, M. V. and Paik, H. J.},
  title = {Superconducting gravity gradiometer for sensitive gravity measurements. II. Experiment},
  journal = {Physical Review D},
  volume = {35},
  pages = {3572},
  year = {1987},
  doi = {10.1103/PhysRevD.35.3572}
}

@article{Griggs2017,
  author = {Griggs, C. E. and Moody, M. V. and Norton, R. S. and Paik, H. J. and Venkateswara, K.},
  title = {Sensitive superconducting gravity gradiometer constructed with levitated test masses},
  journal = {Physical Review Applied},
  volume = {8},
  pages = {064024},
  year = {2017},
  doi = {10.1103/PhysRevApplied.8.064024}
}

@article{Schlamminger2008,
  author  = {Schlamminger, S. and Choi, K.-Y. and Wagner, T. A. and Gundlach, J. H. and Adelberger, E. G.},
  title   = {Test of the equivalence principle using a rotating torsion balance},
  journal = {Phys. Rev. Lett.},
  volume  = {100},
  pages   = {041101},
  year    = {2008},
  doi     = {10.1103/PhysRevLett.100.041101}
}

@article{Wagner2012,
  author  = {Wagner, T. A. and Schlamminger, S. and Gundlach, J. H. and Adelberger, E. G.},
  title   = {Torsion-balance tests of the weak equivalence principle},
  journal = {Class. Quantum Grav.},
  volume  = {29},
  pages   = {184002},
  year    = {2012},
  doi     = {10.1088/0264-9381/29/18/184002}
}

@article{LVKO4a2025,
  author = {{LIGO Scientific Collaboration} and {Virgo Collaboration} and {KAGRA Collaboration}},
  title  = {Direct multi-model dark-matter search with gravitational-wave interferometers
            using data from the first part of the fourth {LIGO}--{Virgo}--{KAGRA} observing run},
  journal = {arXiv e-prints},
  eid = {arXiv:2510.27022},
  year = {2025},
  eprint = {2510.27022},
  eprinttype = {arXiv},
  eprintclass = {astro-ph.CO}
}

@article{companion,
  author  = {Oh, John J. and Paik, Ho Jung},
  title   = {Composition-dependent ultralight dark matter detection with a
             tetrahedral superconducting gravity gradiometer},
  journal = {companion paper, submitted jointly to Phys. Rev. D},
  year    = {2026}
}

@article{Manley2021,
  author = {Manley, Jack and Dey Chowdhury, Mitul and Grin, Daniel and
            Singh, Swati and Wilson, Dalziel J.},
  title = {Searching for Vector Dark Matter with an Optomechanical Accelerometer},
  journal = {Physical Review Letters},
  volume = {126},
  pages = {061301},
  year = {2021},
  doi = {10.1103/PhysRevLett.126.061301}
}

@software{SensitivityCode2026,
  author    = {Oh, John J. and Paik, Ho Jung},
  title     = {Reproduction script for the sensitivity curves of {SGG--ULDM}},
  year      = {2026},
  publisher = {Zenodo},
  doi       = {10.5281/zenodo.22747613},
  url       = {https://doi.org/10.5281/zenodo.22747613},
  note      = {Archived at Zenodo}
}

@incollection{Paik1996,
  author    = {Paik, Ho Jung},
  title     = {Superconducting accelerometers, gravitational-wave transducers,
               and gravity gradiometers},
  booktitle = {SQUID Sensors: Fundamentals, Fabrication and Applications},
  editor    = {Weinstock, H.},
  publisher = {Kluwer Academic Publishers},
  address   = {Dordrecht},
  year      = {1996},
  pages     = {569--598}
}

\section*{End Matter}
\appendix

\section{Dilatonic coupling and the charges of the two materials}
\label{app:charges}

A dilaton-like scalar couples through
\begin{equation}
\begin{aligned}
\mathcal L_{\rm int}=\varphi\bigg[&\frac{d_e}{4e^2}F_{\mu\nu}F^{\mu\nu}
-\frac{d_g\beta_3}{2g_3}G^A_{\mu\nu}G_A^{\mu\nu}\\
&-\sum_i\!\left(d_{m_i}+\gamma_{m_i}d_g\right)m_i\bar\psi_i\psi_i\bigg],
\end{aligned}
\label{eq:Lint}
\end{equation}
with five independent couplings $c\in\{e,g,\hat m,\delta m,m_e\}$ and the field normalized as $\varphi=\kappa\phi$, $\kappa=\sqrt{4\pi G}$. Transferring the field dependence onto the fine-structure constant, the QCD scale and the fermion masses gives~\cite{DamourDonoghue2010,Hees2018}
\begin{equation}
\begin{aligned}
\alpha_A=d_g&+(d_{\hat m}-d_g)Q_{\hat m,A}+(d_{\delta m}-d_g)Q_{\delta m,A}\\
&+(d_{m_e}-d_g)Q_{m_e,A}+d_e Q_{e,A} ,
\end{aligned}
\label{eq:alphaA}
\end{equation}
with the $Q_{c,A}$ the material-dependent dilaton charges. Evaluating the semi-empirical expressions of~\cite{DamourDonoghue2010} in the form tabulated by~\cite{Hees2018} for the two materials gives Table~\ref{tab:charges}, with differences $\lvert\Delta Q'_{\hat m}\rvert=1.910\times10^{-3}$ and $\lvert\Delta Q'_{e}\rvert=1.656\times10^{-3}$. These are values for the pure metals. The gradiometer requires its TMs to be equal in mass. The companion paper~\cite{companion} will show how we construct the TMs of equal mass and equal outer shape although their densities are different.

\begin{table}[t]
\squeezetable
\caption{\label{tab:charges}Dilatonic charges of the two materials. All entries in units of $10^{-3}$.}
\begin{ruledtabular}
\begin{tabular}{lcccc}
& $-Q'_{\hat m}$ & $Q'_{e}$ & $Q'_{\delta m}$ & $-Q'_{m_e}$\\
\colrule
copper & $9.622$ & $2.430$ & $0.1485$ & $0.0240$\\
lead & $7.712$ & $4.086$ & $0.3544$ & $0.0573$\\
\end{tabular}
\end{ruledtabular}
\end{table}

\section{Noise model and the comparison bounds}
\label{app:noise}

The gradient-noise power spectral density, combined over the six differential accelerometer channels of SOGRO, is
\begin{equation}
S_{\Gamma,\Sigma}(f)=\frac{8}{9mL^2}\left[\frac{k_BT\,\omega_D}{Q_D}
+\left(1+\beta^{-2}\right)^{1/2}n\hbar\,\lvert\omega^2-\omega_D^2\rvert\right],
\label{eq:noisemodel}
\end{equation}
a Nyquist thermal noise term and an amplifier noise term of energy resolution $n\hbar$~\cite{Paik2016,Paik2020}. The amplifier term carries the transducer factor $(1+\beta^{-2})^{1/2}$, which the design brings close to unity with $\beta\gg1$ below $1\,$Hz, and the SQUID noise temperature enters through $k_BT_N=n\hbar\omega_p$ at the pump frequency $\omega_p$. The baseline design adopted here is that of mSOGRO with $m=10\,$kg, $L=0.4\,$m, $T=4.2\,$K, $f_D=0.03\,$Hz, $Q_D=10^8$, $n=280$. We assume that this will be upgraded to $m=40\,$kg, $T=0.1\,$K, $Q_D=10^9$, $f_D=10\,$mHz  at a later stage. The sensitivity plotted in Fig.~\ref{fig:sens} is
\begin{equation}
d^{\max}_{\hat m}(f)=\frac{L_{\rm eff}}{A_0\,\lvert\Delta Q'_{\hat m,\rm eff}\rvert}
\sqrt{\frac{2\,S_{\Gamma,\Sigma}(f)}{\tau_{\rm eff}(f)}} ,
\label{eq:sensclosed}
\end{equation}
with $L_{\rm eff}=L/\sqrt2$ the effective baseline of Eq.~(\ref{eq:qresult}), and $\tau_{\rm eff}=\tau_{\rm int}$ below $\tau_{\rm coh}=Q_{\rm coh}/f$ and $\sqrt{\tau_{\rm coh}\tau_{\rm int}}$ above it. Because $A_0$ is independent of $m_\phi$, the shape of each curve is that of the instrument noise itself.

The comparison curves are constructed as
\begin{equation}
\begin{aligned}
d^{\rm stat}_{\hat m}
&=\left[\frac{\lvert\eta\rvert_{95}}
{\bar Q_{\hat m,\oplus}\,\lvert\Delta Q'_{\hat m}\rvert\,S(\lambda)}\right]^{1/2},\\
S(\lambda)&=\frac{3(x\cosh x-\sinh x)}{x^{3}}
\left(1+\frac{r}{\lambda}\right)e^{-r/\lambda} ,
\end{aligned}
\label{eq:microline}
\end{equation}
with $x=R_\oplus/\lambda$, $R_\oplus$ the Earth's radius, $\lambda=\hbar/(m_\phi c)$ and $r$ the source distance. The square root is there because a static test measures a force between two bodies, so the coupling enters twice~\cite{Hees2018}. The inputs are the published $\lvert\eta\rvert_{95}$ of each experiment~\cite{Touboul2022,Schlamminger2008,Wagner2012}, the charges of its TM pair~\cite{Hees2018}, and the Earth as source, treated here as a uniform-density sphere. This gives flat levels of $4.7\times10^{-6}$ for MICROSCOPE and $2.3\times10^{-5}$ for E\"ot-Wash. Both take one coupling nonzero at a time. The sensitivity curves assume that the scalar constitutes the whole of the local dark matter; the static bounds, being sourced by the Earth rather than by the halo, carry no such assumption.

Table~\ref{tab:scaling} records how the sensitivity responds to the design parameters. The array size deserves comment. The signal is independent of $L$, and the gradient noise amplitude $S_{\Gamma,\Sigma}^{1/2}$ of Eq.~(\ref{eq:noisemodel}) also falls as $1/L$, so the two cancel and the coupling sensitivity is independent of $L$ as well. A tidal or gravitational-wave response grows with the baseline instead.

\begin{table}[t]
\caption{\label{tab:scaling}Scaling of the coupling sensitivity of Eq.~(\ref{eq:sens}).}
\begin{ruledtabular}
\begin{tabular}{ll}
Parameter & Dependence of the sensitivity\\
\colrule
charge contrast $\Delta Q'_c$ & $(\Delta Q'_c)^{-1}$\\
test mass $m$ & $m^{-1/2}$\\
quality factor $Q$ & $Q^{-1/2}$\\
temperature $T$ & $T^{1/2}$\\
resonance $\omega_0$ & $\omega_0^{1/2}$\\
integration time & $\tau_{\rm eff}^{-1/2}$\\
array size $L$ & none\\
\end{tabular}
\end{ruledtabular}
\end{table}

\end{document}